\documentclass{sf2a-conf}
\usepackage{graphicx}
\usepackage{natbib}
\begin{document}

\TitreGlobal{SF2A 2008}

%%-----------------------------
%%      the top matter
%%-----------------------------
\title{Rotation on sub-kpc scales in the strongly lensed z$\sim3$ 'arc\&core'
and implications for high-redshift galaxy dynamics}
\author{Nesvadba, N.,~P.,~H.$^{1,}$}
\address{Marie-Curie Fellow}
\address{GEPI, Observatoire de Paris, CNRS,
  Universite Denis Diderot; 5, Place Jules Janssen, 92190 Meudon, France} 
\author{Lehnert,~M.~D.$^2$}
\author{Frye,~B.}
\address{Dublin City University, School of Physics, Glasnevin, Dublin 9, Ireland}
\setcounter{page}{237}

\index{Nesvadba, N.}
\index{Lehnert, M.}

\maketitle

%%-----------------------------
%%      your text
%%-----------------------------
\section{IFU observations of gravitationally lensed and field 
galaxies at z$\sim 2-3$} 
Redshifts z$\sim$2$-$3 represent the cosmologically most important epoch of
star formation and galaxy evolution. Detailed studies of individual galaxies
during this epoch are now possible with integral-field spectrographs
(IFUs). The emerging picture is however far from simple. Even adaptive optics
(AO) assisted observations reach resolutions of only $\sim 1$ kpc, making it
difficult to infer even the basic physical mechanism driving the
kinematics. Forster Schreiber et al. (2006) (hereafter FS06) argue that at
least a subsample of blue, star-forming galaxies at somewhat lower redshifts,
z$\sim 2.5$, may show the signs of large, spatially-extended, rotating disks.
Law et al. (2007) emphasize however that UV selected z$\sim$2$-$3 galaxies
have irregular kinematics, which are likely not dominated by large-scale
gravitational motion, but may be more related to merging or gas cooling.
Moreover, none of these scenarios may be easily generalized to the overall
population of high-z galaxies. To ensure observational success, only bright
and large sources are being targeted with IFUs. This bears the risk that the
targets will not be good representatives of the overall high-z galaxy
population, but may be biased towards the most actively star forming and
disturbed galaxies such as (minor and major) mergers.

To investigate whether this worry is substantiated we compare IFU samples of
z$\sim$2$-$3 galaxies with galaxies at similar redshifts, where the
observational constraints are alleviated by the additional boost of a
gravitational lense. We do this in two steps: (1) Detailed comparison of a
lensed and unlensed z$\sim$ 3 LBG with rest-frame optical IFU data. (2) A
comparison of rest-frame optical line widths in unlensed z$\sim$2$-$3 galaxies
with IFU data and lensed galaxies. Both comparisons suggest that existing IFU
samples may be seriously biased.

\section{The ``arc\&core'': The first z$\sim$3 galaxy with a rotation 
curve on sub-kpc scales} 
Due to a fortuitous lensing configuration of the z$=$3.2 strongly lensed LBG
(SLLBG) 'arc\&core', we see a zoom into the inner $\sim 1$ kpc and several
more peripheral patches magnified by factors $\sim 20$. Most strikingly,
[OIII]$\lambda$5007 line emission reveals a smooth velocity gradient of 190 km
s$^{-1}$ at a spatial resolution of $\sim 200$ pc in the source plane
(Nesvadba et al. 2006), that resembles rotation curves of spiral galaxies at
low redshift (Fig. 1). Line widths are uniform and relatively narrow,
decreasing from $\sigma$$=$97$\pm$9 km s$^{-1}$ in the inner 'core' to
$\sigma$$=$$62\pm15$ km s$^{-1}$ in the outer 'arc'. The overall properties of
the 'arc\&core' appear rather average compared to samples of LBGs in the
field, raising confidence that rotation on sub-kpc scales is not uncommon for
high-z galaxies, and consistent with inside-out disk formation scenarios.
Such scales are significantly smaller than what is found from IFU observations
of unlensed galaxies at slightly lower redshifts.

An important measure to quantify the amount of random to ordered motion in
galaxies is the ratio between velocity gradient $v$ and Gaussian line width
$\sigma$, $v/\sigma$. For the arc\&core, $v/\sigma \sim$3, while the galaxies
of FS06 have $v/\sigma\sim$ 1, suggesting that field galaxies studied with
IFUs are kinematically more strongly disturbed.  This may be a result of
somewhat different selection criteria (the z$\sim 2.5$ sample has a UV color
selection, but not strictly the Lyman-break technique). However, even the
z$=$3.2 LBG Q0347-383 C5 (a classical LBG) shows the same trend. C5 consists
of two separate, unresolved line emitting clumps with a relative velocity
shift of 33 km s$^{-1}$, but much greater line widths $\sigma \sim 85$ km
s$^{-1}$ (Fig. 1), perhaps indicative of a merger of two subunits. The
co-moving space density of bright LBGs like C5 is consistent with
theoretically predicted merger rates (Nesvadba et al. 2008).

A comparison based on only two galaxies cannot be conclusive, but the number
of strongly lensed LBGs at z$\sim$2$-$3 with deep IFU data sets is small, and
the arc\&core is the only SLLBG in the literature with a spatially resolved
rotation curve. We thus compare with the integrated line widths of 5 z$\sim$ 3
SLLBGs with near-IR spectroscopy in the literature. We find significantly more
narrow lines in the integrated spectra of SLLBGs compared to the spatially
resolved maps of FS06. The spatial resolution of the field galaxies
approximately corresponds to the size of the lensed galaxies in the source
plane, so we compare spectra extracted from similarly large regions. Similarly
to the above detailed comparison of two LBGs with IFU data, this may suggest
that great care is warranted when generalizing IFU observations to the overall
population of high-z galaxies.

\begin{figure}[h]
   \centering
   \includegraphics[width=17cm]{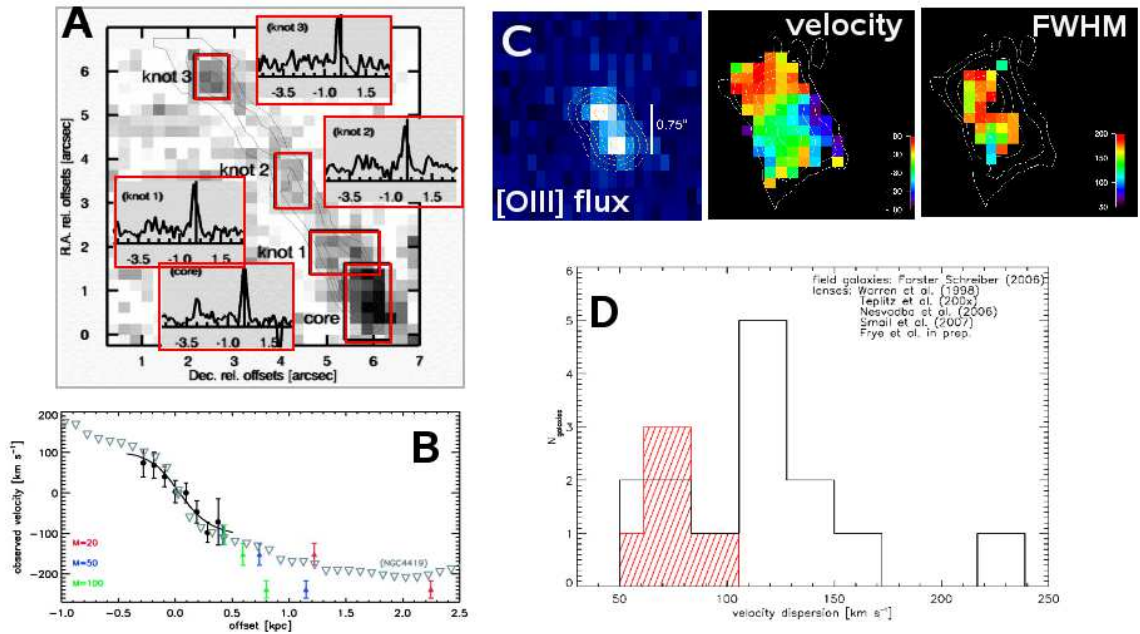}
      \caption{{\it (A) [OIII]$\lambda$5007 line image of the
'arc\&core'. Insets show spectra extracted from the apertures highlighted in
red. (B) Relative velocities within the ``core'' (black dots) and the ``arc''
(red,blue and green triangles) closely resemble the rotation curve of a $\le
10^{10}$ M$_{\odot}$ disk galaxy (black line: model, gray upside-down
triangles: Rotation curve of the ${\cal L}^*$ sprial galaxy NGC4419 in the
Virgo cluster). (C) (left to right) [OIII] morphology, velocities and FWHMs of
the bright, unlensed LBG Q0347-383 C5. Morphology and kinematics are
consistent with two interacting subunits. (D) Comparison of the line widths in
lensed galaxies (red hatched histogram) with the sample of z$\sim$ 2.5
galaxies with IFU data (FS06). The lensed galaxies have widths in the lower
tail of the unlensed sample.}}
\label{fig:arccore}
\end{figure}

{}

%\end{thebibliography}
\end{document}